\documentclass[12pt]{article}
\usepackage{hyperref}
\usepackage{amssymb,amsmath,amsfonts}
\usepackage{color}

\numberwithin{equation}{section}
\DeclareMathAlphabet{\mathcal}{OMS}{cmsy}{b}{n}

\begin{document}

\title{\bf A superspace description of Friedmann-Robertson-Walker models}
 \author{ Sudhaker Upadhyay$^{a}$\thanks {  sudhakerupadhyay@gmail.com; 
 sudhaker@iitk.ac.in}   \\ 
\textit{$^{a}$ \small Department of Physics, Indian Institute of Technology Kanpur,}\\
\textit{ \small Kanpur 208016, India} }
\maketitle
\begin{abstract}
 We illustrate the cosmological Friedmann-Robertson-Walker (FRW) models  realized as gauge theory  
 in the extended configuration space  with its Becchi-Rouet-Stora-Tyutin (BRST) invariance
 upto total derivative. To investigate the model in Batalin-Vilkovisky (BV)
 formalism the Lagrangian density of the models is further extended by introducing shifted fields
 corresponding to all fields.   
   The extended Lagrangian density having shifted fields 
  admits more general BRST symmetry (including shift symmetry), called as extended BRST symmetry. In this framework
  the antighost fields corresponding to the shift symmetry get identified with antifields 
  of BV formulation.   Further, we analyse the  models on supermanifold  with the help of additional super (Grassmannian) coordinate  $\theta$. 
Remarkably, we observe that the $\theta$ components of superfields produces the gauge-fixing term in tandem with ghost term of the effective Lagrangian density naturally.
 Furthermore,
we  show  that the quantum master equation of the BV 
quantization method  can
be translated to  have a superfield structure for the FRW models.
\end{abstract}

\section{Introduction}
Quantum cosmology is a consequence of the efforts toward the development of
a quantum theory of gravity, i.e., unification of quantum mechanics and general relativity  \cite{de,dl}. According to the
cosmological principle,  both the spatial homogeneity and the
isotropy of universe,  which was originally stated for the large scale,  is actually valid
for the very large large scale of the universe. The study about homogeneous and isotropic spacetime symmetry was originally made  by Friedmann, Robertson, and Walker (FRW), see  Refs. \cite{frw, frw1,frw2,frw3,frw4,frw5} and in honour to them such universe models are called as FRW models.
FRW models play a central role in modern cosmology.
Most of the  works on quantum cosmology are based on FRW  universe models although
some authors have studied anisotropic models also (for instance \cite{14}).
 In particular, almost all popular theoretical
models of dark energy get relevance in FRW spacetime. Nevertheless, it is worth mentioning that almost all
the models of dark energy meet
some difficulties like cosmological constant problems, fine-tuning problems and so on. One of
realizations of such difficulties of modern cosmology is the necessity of more careful  investigation of the basics of  FRW cosmology.

The realization of  FRW models as a gauge theory is well established.
When it comes to the quantization of general gauge theories in a Lagrangian formalism, the
BV (also called field/antifield) approach \cite{bv,bv1} appears to be more prevailing and rigorous to all other available schemes.
 In this formulation the solution of a so-called master equation provides
the configuration-space counterpart of the Batalin-Fradkin-Vilkovisky
(BFV) phase-space quantum action \cite{bfv, bfv1, bfv2, bfv3,bfv4}.
The BV formulation encompasses the Faddeev-Popov
quantization and uses the BRST symmetry, which plays a prominent role in the standard paradigm of fundamental interactions, to build on it \cite{brs,brs1}.  The BV formalism, which is based on an action that contains both fields and  antifields, can be thought of as a vast generalization of the original BRST formalism for pure Yang-Mills theory to an arbitrary Lagrangian gauge theory.

 A superspace description for the
non-cosmological modes in BV formulation has been analysed extensively \cite{ad,ad1,ad2, ba,lav, lav1}.
Particularly, Yang-Mills theory \cite{ba}, higher derivative theory \cite{fk},
higher-form gauge theories \cite{subm} have been studied in this context.   Recently, similar 
formulation has been established  for the theory of perturbative gravity at one-loop order
 \cite{sud}. However, for the cosmological models  such analysis has not yet been made.
 Although the BRST symmetry has been analysed for FRW models \cite{mon,jj}.
This provides us motivation to make the analysis in cosmological models.

In this paper, we study the extended BRST symmetry of the FRW models which 
incorporates a shift symmetry in tandem with usual BRST symmetry. Within the
analysis we need ghost and antighost fields corresponding to shift symmetry.
Further we choose a differential gauge fixing
  term to fix the shift symmetry
in such manner that it removes the shift in fields and we recover our original theory.
By doing so, the antighost fields are identified with the antifields of
the BV formulation. Further we discuss the extended BRST invariant models
in superspace.  To make such analysis in superspace we need one more coordinate 
with Grassmannian nature.
Finally, we show that the gauge-fixing and ghost terms of the 
extended BRST invariant models are the outcome of the $\theta$ components of superfields.
The superspace description of quantum master equation at one-loop order for the FRW models is also
analysed. 

This paper is the presented in the following manner. In section 2,  a mathematical formulation of the FRW models is given where the outline of
Hamiltonian dynamics in extended phase space is presented. Further, in Sec. 3, we 
demonstrate the extended FRW models possessing extended BRST transformations (including shift symmetry),  where we derive the
antifields  in more natural way. The extended BRST invariant description of the models on supermanifold  is discussed in Sec. 4.
We discuss the results and future problems in last section.
\section{FRW models and its BRST symmetry}
In this section, we analyse the  BRST symmetry of the cosmological FRW models describing homogeneous and isotropic universe. 
The metric tensor for FRW models in spherical coordinates is given by,
\begin{equation}
ds^2=N^2dt^2+a^2(t)\left(\frac{1}{1-kr^2}dr^2+r^2d\theta^2 +r^2\sin^2\theta  d\phi^2\right),
\end{equation}
where $N$ and $a(t)$ are the lapse function and the scale factor, respectively, depending  on time only and the values of 
$k=1, 0,-1$ correspond  to a space of positive, negative
and zero curvature respectively. 
The classical Lagrangian density of the models  traditionally described in  Arnowitt-Deser-Misner (ADM) variables is given by \cite{mon,tp},
\begin{eqnarray}
{ L}_{inv} =  -\frac{1}{2}\frac{a\dot{a}^2}{N}+\frac{k}{2}Na,\label{cs}
\end{eqnarray}
where parameter $k=1, 0,-1$ refer   a closed, flat and open universe, respectively.
The conjugate momenta corresponding to the lapse function $N$ and the scale factor $a$ read,
\begin{eqnarray}
\pi_N &=&0, \label{pr}\\
\pi_a &=&- \frac{a\dot a}{N},
\end{eqnarray}
where condition (\ref{pr}) describes the primary constraint of the theory.
Using Legendre transform  the  canonical Hamiltonian density is calculated as \cite{mon,tp}
\begin{eqnarray}
H_c =\pi_a\dot{a}-L_{inv} =-\frac{N\pi_a^2}{2a} -\frac{k}{2}Na.
\end{eqnarray}
 Conservation of the primary constraint  (\ref{pr}) with respect to time yields  the   secondary constraint of the theory as follows:
 \begin{equation}
  \frac{\pi_a^2}{2a}+\frac{k}{2}a =0.
 \end{equation}
 Now, it is easy to verify that both the constraints are of first-class (as they commute with each
 other).
 Therefore, we can say with certainty  that the FRW models admit gauge invariance. The  gauge transformation, under which the Lagrangian density (\ref{cs}) remains   invariant,  is given by \cite{mon},
\begin{equation}
\delta N=-N\dot\eta -\dot N\eta,\ \ \ \delta a=-\dot a\eta,\label{gauge}
\end{equation}
 where   $\eta(t)$ is an infinitesimal time dependent parameter of transformation. 
 Before quantizing the theory, it is necessary to impose gauge-fixing condition 
which breaks the  gauge invariance.
This gauge-fixing condition
must satisfy the following requirements: (i) it must fix the
gauge completely, i.e., there must be no residual gauge
freedom (ii) using the transformations   it must be
possible to bring any configuration, specified by $N$ and $a$
into one satisfying the gauge condition.
We choose the   
following gauge condition  satisfying the above mentioned requirements \cite{tp}:
\begin{equation}
\dot{N} =\frac{d}{dt}f(a).\label{gf}
\end{equation}
To employ the  gauge condition (\ref{gf}) in the theory at quantum level we add the
following  gauge-fixing term in the invariant Lagrangian density (\ref{cs}) \cite{mon}:
\begin{eqnarray}
{L}_{gf} &= & \lambda\left( \dot{N}-\frac{d}{dt}f(a)\right), \label{gaf}
\end{eqnarray}
where $\lambda$ is the multiplier (auxiliary) field.

Further, to compensate the effect of above gauge-fixing term from the functional integral, we add the following ghost term   
corresponding to the above gauge-fixing condition in the effective Lagrangian density \cite{mon}:
\begin{eqnarray}
{L}_{gh} &= & \dot{\bar{c}}\left( \dot{N}-\frac{d}{dt}f(a)\right)c +\dot{\bar{c}}N\dot{c},\label{gh}
\end{eqnarray}
where fields $c$ and  $\bar{c}$ are Faddeev-Popov ghost and antighost respectively. 
Now, by adding (\ref{cs}), (\ref{gaf}) and (\ref{gh}) the complete extended Lagrangian density
 reads \cite{mon},
\begin{equation}
{L}_{ext}={L}_{inv} +{L}_{gf} +{L}_{gh}. \label{tot}
\end{equation}
The nilpotent BRST transformations are constructed by replacing the
parameter $\eta$ of (\ref{gauge}) by ghost field $c$ as follows \cite{mon},
\begin{eqnarray}
s_b N &=&-(\dot{N}  c+N\dot c),\nonumber\\
s_b a &=& -\dot{a} c,\nonumber\\
s_b  c&=& 0,\nonumber\\
s_b\bar c&=&-\lambda,\nonumber\\
s_b \lambda &=&0,
\end{eqnarray}
under which the extended Lagrangian density ${L}_{ext}$ is invariant. 
Since the combination ${L}_{gf} +{L}_{gh}$ is
BRST exact and, therefore, we can express it in terms of BRST variation of gauge-fixing fermion
$\Psi$ as follows,
\begin{equation}
{L}_{gf} +{L}_{gh}= s_b \Psi= -s_b\left[\bar c\left( \dot{N}-\frac{d}{dt}f(a)\right)\right],\label{ps}
\end{equation}
where the gauge-fixing fermion is defined as $\Psi =-\bar c\left( \dot{N}-\frac{d}{dt}f(a)\right)$.
\section{Extended BRST invariant FRW model}
In this section, we analyse the extended BRST transformations for FRW  models.  To do so, let us start  by shifting the fields from their original values
as follows \cite{ba},
\begin{equation}
N \longrightarrow  N- \tilde N \quad
a \longrightarrow  a - \tilde a \quad
\bar c \longrightarrow \bar c - \tilde {\bar c} \quad
c \longrightarrow c- \tilde c\quad
\lambda \longrightarrow \lambda- \tilde \lambda.\label{brs}
\end{equation}
With these shifts in fields, the extended Lagrangian density (\ref{tot}) is given by
\begin{eqnarray}
\tilde{L }_{ext}&= & {L}_{ext}(N- \tilde N, a - \tilde a, \bar c - \tilde {\bar c}, c- \tilde c, \lambda- \tilde \lambda),\nonumber\\
 &= &  -\frac{1}{2}(a-\tilde{a})\frac{(\dot{a}-\dot{\tilde{a}})^2}{N-\tilde{N}}+\frac{k}{2}(N-\tilde N) (a-\tilde{a}) +(\lambda-\tilde{\lambda})\left( \dot{N}-\dot{\tilde{N}}-\frac{d}{dt}f(a-\tilde{a})\right)\nonumber\\
 &+& (\dot{\bar{c}}-\dot{\tilde{\bar{c}}})\left( \dot{N}-\dot{\tilde{N}}-\frac{d}{dt}f(a-\tilde{a})\right)(c-\tilde{c}) +(\dot{\bar{c}}-\dot{\tilde{\bar{c}}})(N-\tilde{N})
 (\dot{c}-\dot{\tilde{c}}).\label{til}
\end{eqnarray}
This extended Lagrangian density is invariant under the same BRST structure (\ref{brs}) but for shifted 
fields.
Here, we notice that the extended Lagrangian density is also invariant under the following shift symmetry  \cite{ba}
\begin{eqnarray}
s  \Phi (x)= \alpha (x),\ \ s  \tilde\Phi (x)&=& \alpha (x),
\end{eqnarray}
where $\Phi$ and $\tilde \Phi$ denote the original fields and the shift in fields   collectively.
 The BRST symmetry together with the shift symmetry manifests  the extended BRST symmetry.
Hence, the extended BRST transformation  can be  defined, compactly,   as  \cite{ba}
\begin{eqnarray}
s_b \Phi (x)= \alpha (x),\ \ s_b \bar\Phi (x)&=& \alpha (x)-\beta (x).
\end{eqnarray}
Here $\beta (x)$ refers the original BRST variation,
whereas $\alpha (x)$ denotes the change under shift symmetry.
 
The  extended BRST symmetry transformation, under which the Lagrangian density (\ref{til}) remains invariant, is constructed explicitly as
\begin{eqnarray}
s_b N &=& \zeta,\ \ s_b \tilde N=\zeta+(\dot{N}  c- \dot{\tilde{N}}  \tilde c+ N\dot c-\tilde N\dot{\tilde c}),\nonumber\\
s_b a &=& \varepsilon,\ \ \ s_b \tilde a = \varepsilon +\dot{a} c -\dot{\tilde{a}}\tilde c\nonumber\\
s_b  c&=&  \vartheta,\ \ \ s_b  \tilde c=  \vartheta,\nonumber\\
s_b\bar c&=&\varsigma,\ \ \ s_b\tilde{\bar c}= \varsigma+\lambda -\tilde{\lambda},\nonumber\\
s_b \lambda &=&\varrho, \ \ \ s_b \tilde\lambda =\varrho,\label{ex}
\end{eqnarray}
where $\zeta, \varepsilon, \vartheta, \varsigma$ and $\varrho$ are the introduced ghost fields 
for the fields $N, a, c, \bar c$ and $\lambda$, respectively, having following ghost numbers:
\begin{eqnarray}
\mbox{gh}(\zeta)=1,\ \ \mbox{gh}(\varepsilon)=1,\ \ \mbox{gh}(\vartheta)=2,\ \ \mbox{gh}(\varsigma)=0,\ \ \mbox{gh}(\varrho)=1.
\end{eqnarray} 
The nilpotency property of  extended BRST symmetry transformations (\ref{ex}) restrict   
the ghost fields $\zeta, \varepsilon, \vartheta, \varsigma$ and $\varrho$  
in such manner that these vanish under BRST transformation, i.e. 
\begin{eqnarray}
s_b \zeta &=& 0 , \nonumber\\
s_b \varepsilon &=& 0 , \nonumber\\
s_b \vartheta &=& 0 , \nonumber\\
s_b \varsigma  &=& 0, \nonumber\\
s_b \varrho  &=& 0.
\end{eqnarray}
However, the  requirement of the physical theory is that the ghost number og the Lagrangian density must be zero .  So, to make the theory   ghost free we 
incorporate the following antighost fields $N^\star_\mu, a^\star, \bar c^\star, c^\star$ and 
$\lambda^\star$ having ghost numbers opposite to that of the respective ghost fields.  Now, we propose that these antighost fields transform under the  BRST transformation as follows, 
\begin{eqnarray}
s_b N^\star  &=& -K, \nonumber\\
s_b a^\star &=& -l, \nonumber\\
s_b \bar c^\star &=& -\bar m, \nonumber\\
s_b c^\star &=& -m,\nonumber\\
s_b \lambda^\star &=&-n,\label{br}
\end{eqnarray} 
where $K, l, n$ are the bosonic Nakanishi-Lautrup type auxiliary fields and $\bar m, m$ are
 fermionic auxiliary fields.
The nilpotency property of BRST transformation (\ref{br}) again reflects
\begin{eqnarray}
s_b K &=& 0 , \ \ s_b l= 0 , \ \
s_b \bar m = 0 , \ \ s_b m= 0, \ \ s_b n=0.
\end{eqnarray}
Now, one the possible way  to recover the original FRW models described by Lagrangian density
 (\ref{cs}) is to make a suitable gauge condition which fixes the shift 
symmetry such that all the tilde fields vanish.
The suitable  gauge-fixing term 
which fixes the shift symmetry   is constructed as follows,
\begin{eqnarray}
\tilde{L}_{gf }+\tilde{L}_{gh } &=&   -K \tilde N - N^\star [\zeta+(\dot{N}  c- \dot{\tilde{N}}  \tilde c+ N\dot c-\tilde N\dot{\tilde c})]-l\tilde a -a^\star (\varepsilon +\dot{a} c -\dot{\tilde{a}}\tilde c) - m \tilde {\bar c }
 \nonumber\\
& 
+& c^\star[\varsigma+\lambda -\tilde{\lambda}]-\bar m \tilde c + \bar c^\star \vartheta -n \tilde \lambda - \lambda^\star \varrho.\label{ij}
\end{eqnarray}
After performing  integration  over the auxiliary fields (within functional integral) 
the above expression (\ref{ij}) reduces to
\begin{eqnarray}
\tilde{L}_{gf }+\tilde{L}_{gh } &=&   - N^\star [\zeta+(\dot{N}  c + N\dot c )]  -a^\star (\varepsilon +\dot{a} c  )  
 \nonumber\\
& 
+& c^\star[\varsigma+\lambda] + \bar c^\star \vartheta  - \lambda^\star \varrho .\label{i}
\end{eqnarray}
The original gauge-fixing and ghost terms of the model in terms of general gauge-fixing 
fermion $\Psi$  
can be described by, 
\begin{eqnarray}
{ L}_{gf}+{ L}_{gh} &= &  s_b \Psi = \left[s_b N \frac{\delta \Psi}{\delta N} + s_b a \frac{\delta \Psi}{\delta a} 
+ s_b \bar c \frac{\delta \Psi}{\delta \bar c} +  s_b c \frac{\delta \Psi}{\delta c}+  s_b \lambda \frac{\delta \Psi}{\delta \lambda}\right], \nonumber\\
&= &\left[\zeta\frac{\delta \Psi}{\delta N} + \varepsilon \frac{\delta \Psi}{\delta a} 
+ \varsigma\frac{\delta \Psi}{\delta \bar c} +  \vartheta \frac{\delta \Psi}{\delta c}+ \varrho \frac{\delta \Psi}{\delta \lambda}\right].
\label{g}
\end{eqnarray}
Now, the complete  gauge-fixing and ghost terms (sum of  (\ref{i}) and (\ref{g})) which 
fix the extended BRST symmetry are given by,
\begin{eqnarray}
  \tilde { L}_{gf} + \tilde { L}_{gh}+{L}_{gf} 
+ { L}_{gh}
  &=&  \left(- N^\star - \frac{\delta \Psi}{\delta N} \right)\zeta  +
  \left(-a^\star - \frac{\delta \Psi}{\delta a}\right) \varepsilon +
  \left(c^\star + \frac{\delta \Psi}{\delta \bar c}\right)\varsigma  \nonumber\\
 &+&   \left(\bar c^\star +\frac{\delta \Psi}{\delta c}\right) \vartheta -
  \left(\lambda^\star + \frac{\delta \Psi}{\delta \lambda}\right)\varrho
- N^\star (\dot{N}  c + N\dot c ) \nonumber\\
& -& a^\star  \dot{a} c     
  + c^\star \lambda.\label{eff}
\end{eqnarray}
If we  integrate out the ghost fields $\zeta, \varepsilon, \varsigma, \vartheta$ and $\varrho$ associated with the shift
symmetry, the above expression rests with
\begin{eqnarray}
  \tilde { L}_{gf} + \tilde { L}_{gh}+{L}_{gf} 
+ { L}_{gh}
  &=&  
- N^\star (\dot{N}  c + N\dot c ) - a^\star  \dot{a} c     
  + c^\star \lambda,\label{eff1}
\end{eqnarray}
leading to following constraints:
\begin{eqnarray}
N^\star   &=& -\frac{\delta \Psi}{\delta N}, \ \ a^\star =- \frac{\delta \Psi}{\delta a}  , \nonumber\\
c^\star &=& - \frac{\delta \Psi}{\delta \bar c}  , \ \
\bar c^\star = -\frac{\delta \Psi}{\delta c},\nonumber\\
\lambda^\star &=& -\frac{\delta \Psi}{\delta \lambda}. \label{ant}
\end{eqnarray}
Here we observe that the antighosts of the theory are identified with the antifields of BV formulation. The consistency of the result can be checked  as follows:
for the theory of FRW model the  expression for gauge-fixing fermion $\Psi$ is given in (\ref{ps}).
For that $\Psi$  the  antighost fields established in (\ref{ant}) get following identification  
\begin{eqnarray}
N^\star   &=&  -\dot{\bar c}, \nonumber\\
 a^\star &=&\dot{\bar c}\frac{d}{da}f(a)  , \nonumber\\
c^\star &=&  \left(\dot{N}- \frac{d}{dt}f(a)\right)  ,\nonumber\\
\lambda^\star &=& 0, \ \
\bar c^\star = 0. 
\end{eqnarray}
Plugging back these specific values of antighosts in (\ref{eff1}), we recover the sum of our 
original gauge-fixing and ghost terms of FRW models given in (\ref{gaf}) and (\ref{gh}).
Remarkably, we noticed that the original FRW theory is recovered from the
models in extended configuration space possessing extended BRST symmetry where antighosts are 
identified 
with antifields (of BV formalism) naturally.
 
\section{FRW model in superspace}
In this section, we analyse the extended BRST invariant FRW models on the five dimensional 
supermanifold.
To describe the model in superspace, we first define  the 
 superfields, which depend on five super-coordinates $(x,\theta)$ such that at vanishing $\theta$ these superfields identified with original fields,  as follows
\begin{eqnarray}
{\cal N} (x,\theta ) &=& N(x) + \theta \zeta , \nonumber\\
\tilde {\cal N} (x,\theta ) &=& \tilde N(x) + \theta [\zeta+(\dot{N}  c- \dot{\tilde{N}}  \tilde c+ 
N\dot c-\tilde N\dot{\tilde c})] , \nonumber\\
{\cal A} (x,\theta ) &=& a(x)+ \theta \varepsilon , \nonumber\\
\tilde {\cal A} (x,\theta )&=& \tilde a(x) + \theta [ \varepsilon +\dot{a} c -\dot{\tilde{a}}\tilde 
c] , \nonumber\\
{\cal C} (x,\theta ) &=& c(x) + \theta  \vartheta , \nonumber\\
\tilde {\cal C} (x,\theta ) &=& \tilde c(x) + \theta \vartheta , \nonumber\\
\bar{\cal C} (x,\theta )  &=&\bar c(x) + \theta \varsigma , \nonumber\\
\tilde{\bar{\cal C}} (x,\theta )  &=& \tilde {\bar c}(x) + \theta [\varsigma+\lambda -\tilde{\lambda}],\nonumber\\
  {  \Lambda}(x,\theta )  &=& \lambda(x)+ \theta \varrho , \nonumber\\
\tilde{   \Lambda} (x,\theta )  &=& \tilde \lambda (x)+ \theta \varrho,\label{sup}
\end{eqnarray}
 where $\theta$ is Grassmannian coordinate.
Similarly, we define the super-antifields which depend  on super-coordinates 
$(x,\theta)$ whose vanishing $\theta$ components yields original four dimensional local antifields,
\begin{eqnarray}
{\cal N}^\star (x,\theta ) &=& N^\star(x) - \theta K, \nonumber\\
{\cal A}^\star (x,\theta ) &=& a^\star(x)-\theta l, \nonumber\\
{\cal C}^\star (x,\theta ) &=& c^\star(x) - \theta  m, \nonumber\\
\bar{\cal C}^\star (x,\theta )  &=&\bar c^\star(x) - \theta \bar m, \nonumber\\
  {  \Lambda}^\star(x,\theta )  &=& \lambda^\star(x)- \theta n.\label{asup}
\end{eqnarray}
From the above  expressions of  superfields and super-antifields given in (\ref{sup}) and (\ref{asup}), respectively, we  compute the following relations
\begin{eqnarray}
\frac{\delta}{\delta \theta}({\cal N}^\star \tilde {\cal N}) &=& -K \tilde N - N^\star [\zeta+(\dot{N}  c- \dot{\tilde{N}}  \tilde c+ N\dot c-\tilde N\dot{\tilde c})], \nonumber\\
\frac{\delta }{\delta \theta}({\cal A}^\star \tilde {\cal A}) &=&-l\tilde a -a^\star (\varepsilon +\dot{a} c -\dot{\tilde{a}}\tilde c), \nonumber\\
\frac{\delta }{\delta \theta}(\tilde {\bar{\cal C}} {\cal C}^\star) &=& - m \tilde {\bar c }
+c^\star[\varsigma+\lambda -\tilde{\lambda}], \nonumber\\
\frac{\delta }{\delta \theta}(  {\bar{\cal C}^\star} \tilde{\cal C}) &=&- \bar m \tilde c + \bar c^\star \vartheta, \nonumber\\
\frac{\delta }{\delta \theta}(\Lambda^\star \tilde \Lambda) &=&-n \tilde \lambda - \lambda^\star \varrho, \label{q}
\end{eqnarray}
where derivatives with respect to $\theta$ are considered from left side. Now, from the expression (\ref{q}) it is easy to derive  
   the following relation
\begin{eqnarray}
\frac{\delta}{\delta \theta} ({\cal N}^\star \tilde {\cal N}+{\cal A}^\star \tilde {\cal A}+\tilde {\bar{\cal C}} {\cal C}^\star + {\bar{\cal C}^\star} \tilde{\cal C}+\Lambda^\star \tilde \Lambda)&=& -K \tilde N - N^\star [\zeta+(\dot{N}  c- \dot{\tilde{N}}  \tilde c+ N\dot c-\tilde N\dot{\tilde c})]-l\tilde a \nonumber\\
&-&a^\star (\varepsilon +\dot{a} c -\dot{\tilde{a}}\tilde c)  - m \tilde {\bar c }
+c^\star[\varsigma+\lambda -\tilde{\lambda}]- \bar m \tilde c  \nonumber\\
&+& \bar c^\star \vartheta -\tilde \lambda - \lambda^\star \varrho.\label{ga}
\end{eqnarray}
Here we observe that the r.h.s. of above relation  exactly coincides with the shifted gauge-fixed Lagrangian density  given in (\ref{ij}).
Further, we define the  gauge-fixing fermion on supermanifold    
\begin{eqnarray}
\Gamma(x, \theta) = \Psi (x)+ \theta (s_b\Psi).\label{bs}
\end{eqnarray}
The expression (\ref{bs}) suggests that the original gauge-fixing and ghost  Lagrangian densities 
for FRW models can be acquired from the left derivation of gauge-fixing fermion in superspace with 
respect to $\theta$ as follows:
\begin{eqnarray}
{L}_{gf}+{L}_{gh}=  \frac{\delta \Gamma(x, \theta) }{\delta \theta}.\label{gau}
\end{eqnarray}
From relations (\ref{ga}) and (\ref{gau}), we therefore  conclude that  the complete
gauge-fixed Lagrangian density of extended BRST invariant FRW models can simply
be derived from the $\theta$ components as follows
\begin{eqnarray}
{L}_{gf}+{L}_{gh}+\tilde{L}_{gf}+\tilde{L}_{gh} &=&  \frac{\delta}{\delta \theta} ({\cal N}^\star \tilde {\cal N}+{\cal A}^\star \tilde {\cal A}+\tilde {\bar{\cal C}} {\cal C}^\star + {\bar{\cal C}^\star} \tilde{\cal C}+\Lambda^\star \tilde \Lambda +\Gamma).
\end{eqnarray}
Being $\theta$ component of superfields it is obvious that the complete
gauge-fixed Lagrangian density of extended BRST invariant FRW models in superspace remains 
intact under extended BRST transformations.

Furthermore, to investigate the BRST variation of the quantum action in the standard BV
quantization method,  we first define the
vacuum functional for FRW models
 in BV formulation as
\begin{equation}
Z_\Psi = \int \prod D\Phi \exp \left[\frac{i}{\hbar} W\left( \Phi, \Phi^\star = 
\frac{\partial \Psi}{\partial \Phi}\right) \right], 
\end{equation}
where $\Phi$ and  $\Phi^\star$ are the generic fields
and corresponding antifields of 
the theory. However, $W$ is an
 extended quantum action  of the theory.
 
 The condition of gauge independence of generating functional, which
  translates into  so-called quantum master equation, is given by \cite{sud}
 \begin{equation}
 \frac{1}{2}\left(\frac{\partial_r W}{\partial \Phi}\frac{\partial_l W}{\partial
  \Phi^\star}-\frac{\partial_r W}{\partial \Phi^\star}\frac{\partial_l W}{\partial
  \Phi }\right) = i\hbar \Delta W,\label{bra}
 \end{equation}
 where the operator $\Delta$ is
 defined by
 \begin{eqnarray}
    \Delta = \frac{\partial_r }{\partial\Phi} \frac{\partial_l }
  {\partial\Phi^\star}.
  \end{eqnarray}
   The extended quantum action can be written upto the one-loop order correction
   by
   \begin{equation}
   W(\Phi, \Phi^\star )=S_{ext}(\Phi, \Phi^\star ) +\hbar M_1(\Phi, \Phi^\star ),
   \end{equation}
where $S_{ext}$ is the  action corresponding to (\ref{tot}) and $M_1$ appears from nontrivial measure factors.    
      
The gauge theory having no anomaly upto first-order correction $M_1$ 
   does not depend on the antifields and, consequently, the BRST transformations of the  
   action $S_{ext}$ and $M_1$
   are given by
   \begin{equation}
   s_b S_{ext}=0, \ \ s_b M_1 =i\Delta S_{ext}.
   \end{equation}
    Now, we apply the $\Delta$ operator on total action $S_T$ 
    (having both the original and shifted actions) as
    \begin{equation}
    \Delta S_{ext}=\Delta S_T =\frac{\partial_r }{\partial\Phi} \frac{\partial_l }{\partial\Phi^\star}S_T.
    \end{equation}
Here we note that generic fields $\Phi$ and $\Phi^\star$ includes all the fields, shifted fields,
 ghosts and   and corresponding antighosts fields. 
   Therefore, at the one-loop order, it is logical to define a superfield as
   \begin{equation}
   {\cal M}_1  (x, \theta)= M_1 (x) +\theta i\Delta S_T.
   \end{equation}
However, in the superspace,  the extended quantum action is described by
\begin{equation}
{\cal W} (x, \theta)=W(x) +\theta i\hbar \Delta W.
\end{equation}
  Therefore, the quantum master equation for FRW models 
  can simply be derived from the above relation as following
  \begin{eqnarray}
  \frac{\partial}{\partial \theta }{\cal W}=i\hbar \tilde \Delta {\cal W},
  \end{eqnarray}
  where $\tilde\Delta$ operator 
has the following expression:
\begin{eqnarray}
   \tilde \Delta = \frac{\partial_r }{\partial\Phi (x,\theta)} \frac{\partial_l }
  {\partial\Phi^\star(x,\theta )}.
  \end{eqnarray}
 Therefore, by enlarging the configuration space with the variable $\theta$, the quantum master equation is equipped  with Grassmannian translations that reproduce the effect of the antibrackets
given in (\ref{bra}).
\section{Conclusion}
In modern cosmology the FRW models play a central role.
These cosmological models assume zero cosmological constant. The only force acting is gravity. 
 In this paper we have considered FRW models describing 
 closed, flat and open universe. Further, we have demonstrated the BRST symmetry of the 
 models and have analysed the extended BRST symmetry 
 in which we have made a linear shift in all the fields.
We have recovered the original theory from the shifted Lagrangian density by adding a suitable
gauge-fixing term. The BV procedure represents
a very powerful framework for the quantization of general gauge theories.
The advantage of extending the
phase space by shifting the fields is that the antifields of BV formulation get their identifications naturally.
Furthermore, it is worth to analyse the model on supermanifold.
We have described the models on five dimensional superspace with coordinates $(x, \theta)$. 
For a general gauge fixing fermion, we have shown that the shifted Lagrangian density
 can be written in a
manifestly extended BRST invariant manner in a superspace with one Grassmann coordinate.
 We have shown that the quantum master equation of the standard BV formalism can be represented as 
 the requirement of a superspace structure for the extended quantum action.
 We hope this formulation will be helpful in explaining the FRW models systematically.

It would be interesting to develop anti-BRST transformation for the FRW models. 
With the help of anti-BRST transformation (where the role of ghosts and antighosts are interchanged
with some coefficients), 
we can analyse the models on
six dimensional superspace. However, without 
anti-BRST transformation, we are constrained to
analyse  the models upto five dimensional superspace. 

\end{document}